# Decentralized Lifetime Minimizing Tree for Data Aggregation in Wireless Sensor Networks

Deepali Virmani , Satbir Jain

*Abstract* — To meet the demands of wireless sensor networks (WSNs) where data are usually aggregated at a single source prior to transmitting to any distant user, there is a need to establish a tree structure inside any given event region. In this paper , we propose a novel technique to create one such tree, which preserves the energy and minimizes the lifetime of event sources while they are constantly transmitting for data aggregation in future WSNs. We use the term Decentralized Lifetime-Minimizing Tree (DLMT) to denote this tree. DLMT features in nodes with higher energy tend to be chosen as data aggregating parents so that the time to detect the first broken tree link can be extended and less energy is involved in tree maintenance. In addition, by constructing the tree in such a way, the protocol is also able to reduce the frequency of tree reconstruction, minimizes the amount of data loss ,minimizes the delay during data collection and preserves the energy . Forwarded directed Diffusion protocol is chosen as the routing platform.

## I. INTRODUCTION

Wireless Sensor Networks may deploy several hundreds to thousands of sensor nodes. Protocols in such networks must therefore be scalable. Furthermore, since nodes are untethered and their geographic positions are not pre-determined, these nodes may also need to possess some self organizing capabilities. Network dynamics that result from both node movement and unpredictable energy depletion also bring new challenges to the design of an efficient WSN. Since nodes can only carry limited battery resources, they usually get disconnected from the network easily. Such frequent node disconnections suggest that the design must accommodate topological changes. Since sensors are being densely-deployed in WSNs, the detection of a particular stimulus can trigger the response from many nearby nodes. Thus, data in such networks are usually not directly transmitted to interested users upon event detection. Instead, they are aggregated with neighboring sources locally to remove any redundancy and produce a more concrete reading[1][2][3][4] . In this paper, we focus on constructing a data aggregation tree among any given set of source nodes. The tree has a dedicated root for which the data from various sources are gathered. Moreover, the tree is structured in a way that can preserve the functional node lifetime of the event sources subject to the condition that they are constantly transmitting. The functional node lifetime is defined as the time till a node runs out of its energy. Reference[5] suggests that extending the node lifetime is equivalent to increasing the amount of information gathered by the tree root when the data rate is not time-varying. We consider a network of randomly-deployed sensor nodes in which each node has an identical transmission range. An event that triggers the sensors around it occurs at random in the network. Data reports from these sensors are clock-driven upon event detection. Furthermore, they are aggregated along their ways to be collected at the tree root and periodically sent to the sinks. To prevent data lost, the tree is periodically scanned and any broken link should be repaired whenever necessary. Such tree allows all raw data reports to be aggregated along the way to a single processing point. Only relevant information is extracted before transmitting it to any distant sink. Therefore, the converged tree construction becomes one of the fundamental issues for aggregation in WSNs. In fact, not all the trees are ideal for aggregation inside the event region. Since energy is usually scarce in WSNs, it is most power-efficient if these sources can provide data to the sinks for the longest possible time. A tree that can survive for longer duration thus naturally becomes the best choice. The time till the first link breaks in a given tree structure determines the lifetime of each source and the term tree energy directly reflects this time .We hence tackle this problem by searching a tree that comprises the highest tree energy .

## II. PROBLEM FORMULATION

The conventional spanning tree[12] fails to consider residual energy of nodes in the tree constructions. There is thus a good possibility that a low-energy node is arranged to forward data for some other nodes, thereby reducing its node lifetime and fastening the energy-depletion of any subsequent event source E-Span improves the design of tree construction by assigning root to be the highest energy node. Such arrangement provides root with the maximum amount of energy resources for its additional duty in coordinating the route to distant sinks. However, there is still a high chance of assigning low-energy nodes to be the data aggregating agents for the other sources. To shorten the time and minimize the energy cost to tree reconstructions, and hence preserve the functional lifetime of all sources, we have proposed a Decentralized lifetime-minimizing tree construction algorithm which arranges all nodes in a way that each parent will have the maximal-available energy resources to receive data from all of its children. Such arrangement extends the time to refresh the tree and lowers the amount of data lost due to a broken tree link before the tree reconstructions In the literature, network lifetime has often been defined as either the time till the first or a set of nodes runs out of its energy [5, 6, 7, 8], or till the first loss of connectivity or coverage [9,10], or a combination of these [11]. So network lifetime is interpreted as the time before the network ceases to provide the type of service it is designed for. We therefore follow this convention, and define the branch energy as the minimum energy of all the non-leaf nodes in a given branch and define tree energy as the minimum branch energy of all the branches in a given tree.

Let

$brE_{x,y,A}$ : Energy of branch A leafed at node x and rooted at node y, $A \varepsilon P_{x,y}$.
$TE_i$ : Energy of tree rooted at node i

Branch energy and Tree energy is calculated as :

$$brE_{x,y,A} = \min_{i \in A, i \neq A} \{e_i\} \quad (1)$$

$$TE_i = \min_{j \in i, j \neq x} \{e_j\} \quad (2)$$

The time for an upstream link along a given branch to break directly depends on the energy of the parent on such a link. In other words, the time during which data from each source along this branch can arrive at the root will depend on the minimum energy of any parent along this branch. By using the same analogy, the time during which data from all sources can arrive at the root without having to concern about broken link repairs and tree reconstructions will depend on the minimum energy of any branch, or equivalently that of any parent, in a given tree. The only question we are left with lies on how to select an appropriate tree root and the branch leading to each other source, such that the energy of this tree is maximized.

To resolve this problem, we explore the highest-energy branch from each source to a root by first assuming that every source node is a root. This generates a total of *N* unique trees with residual energy $e_n$ and each being rooted at a distinct source node. $P_{s,t}$ is the set of possible routes with each labeling p , from nodes s to t .We continue by comparing the energy of these trees and only employ the one with the highest tree energy for data collection /aggregation . Our DLMT construction problem is thus formulated as follows:

---

**Construct a tree rooted at node z such that**

1) $TE_z \geq TE_n$ . $\forall n \in N, n \neq z$
   subject to condition that

2) $brE_{x,z,A} \geq brE_{x,z,p}$    $\forall i \in N$, $\forall p \in P_{i,z}$, $p \neq A$

---

III. LIFETIME MINIMIZING TREE : A DISTRIBUTED APPROACH

Given a number of N source nodes with each node labeling i (i ε { 1,2,3…….N } ) and the residual energy $e_i$ .Our goal is to construct a tree spanning all these sources and select an appropriate root for data collection, in a distributed way, such that the energy of the tree is maximized. We take the approach of exploring the highest-energy branch from each source to a root, by first assuming that every source node is a root, using a method similar to Reverse-Path Forwarding (RPF) [13]. This generates a total of *N* unique trees with each being rooted at a distinct source node. We continue by comparing the energy of these trees and only employ the one with the highest tree energy for data collection /aggregation . First of all we describe the procedure to explore the highest energy branch among all the sources in a given tree root .Next we construct N trees with each tree rooted at a distant source by incrementally attaching any tree branch form the previous step , after that we compare these trees and employ the one with the highest tree energy for data aggregation/ collection and at last we present a concrete algorithm of DLMT for practical implementation in wireless sensor networks .

A. *Exploring the Highest Energy Branch from every source to any Root*

The time during which data from each source along a given branch can actively be received by the root depends on the minimum energy of any parent along this branch. In order to maximize this time for any pair of root and source, the connectivity between them will first have to be explored prior to getting the highest-energy branch connecting these two nodes . Let $PR_{a,b}$ denotes the set of possible routes with each labeling p from nodes a to node b and $BrE_{x,y,A}$ be the Energy of branch A leafed at node x and rooted at node y such that A ε $P_{x,y}$. We therefore want to get to a branch b for every pair of source s ant root r such that :

$$brE_{s,r,b} \geq brE_{s,r,p} \quad \forall p \in P_{s,r} \quad , p \neq b \quad (3)$$

this is done by implementing RPF where each source s initiates a configuration message that contains its energy information . When a source receives this message it appends its energy information and broadcasts this message only if it has previously forwarded message contains lower energy branch, else the message is discarded. By this manner message will traverse through various different routes and only the better one will arrive at root r . we define $eid_n$ to be the pair of energy level and ID be the label of node n and $brlist_{a,b,M}$ be the list containing the eid for the message initiated from node a up to the last receiving node b via a route M with branch energy $brE_{a,b,M}$ calculated using (1) .

Format of $brlist_{a,b,M}$ is as follows :

$$\text{brlist}_{a,b,M} : eid_a \to eid_x \to eid_y \to eid_b \quad (4)$$

where x,y are the intermediate nodes receiving the message . The procedure of exploring the *Highest Energy Branch* function is described below:

---

*Highest Energy Branch* Function(node ID i , node energy $e_i$ )

1. Create $brlist_{i,i,-}$ by appending $eid_i$ and single–hop broadcast $brlist_{i,i,-}$
2. while receiving $brlist_{a,b,M}$ from node b (M ε $P_{a,b}$, a ε N , a ≠ i ) ,
3. if node i has not seen the message initiating node a ,
4. append $eid_i$ to the head of $brlist_{a,i,p}$ and update $brE_{a,i,p}$ (p ε $P_{a,i}$ )
5. store and single-hop broadcast $brlist_{a,i,p}$
6. else if min { ei , $brE_{a,b,M}$ } > stored $brE_{a,i,q}$ (q ε $P_{a,i}$ ) ,
7. remove the stored $brlist_{a,i,q}$ and $brE_{a,i,q}$
8. append $eid_i$ to the head of $brlist_{a,b,M}$ and update $brE_{a,i,p}$ ( p ε $P_{a,i}$ )
9. store and single-hop broadcast $brlist_{a,i,p}$ .

---

In the above function single-hop broadcast refers to the operation of sending a packet to all single-hop neighbors .Line 1 allows each source to initiate its control message. Lines 2 to 5 update, store and broadcast the message when receiving node does not recognize the initiating source. Lines 6 to 9 reset its stored list in addition to the above three actions whenever the receiving node detects a higher energy branch.

*B. Construction of a tree that spans all event nodes for every source*

We now proceed to construct N trees with each tree rooted at a distant source by incrementally attaching any branch explored in the last section. Each source has an initial tree structure that only comprises the node itself. In order to construct a tree for each source that spans all event nodes, each source has to incrementally update its existing tree structure upon receiving any branch with an unknown initiating node. Note that the energy of the received branch directly determines the energy of the updated tree. To ensure that each tree carries the highest energy, the tree is also updated whenever the receiving node identifies a message with a higher branch energy. Example of final tree construction for the node a is shown in the Fig. 1a. Fig.1b shows the selection of highest energy branch for various nodes . Besides attaching new branches each source is also responsible for preserving the loop free property of its tree during the update .So the branch that violates this property should be rejected . As shown in Fig. 2 message initiated by node g is first received by node a via node c , but actually message initiated by node g was transmitted to node a via node b .So a loop around nodes a,b,c,g is created . Therefore node g will spend twice its energy to transmit and receive any data generated by it .So in order to remove the loop and to reduce its energy usage each node always has to reject a branch when the already attached ones for each parent on this branch does not match with it .

Let initiator be the source which initiates the message and $j_i$ denote the set of initiators in $tree_j$ and $TE_k$ be the energy of $tree_K$. $TE_k$ can be calculated by using the equation

$$TE_K = \min_{j \in j_K} \{ brE_{j,K,p} \} \quad \text{where } p \in P_{j,K} \quad (5)$$

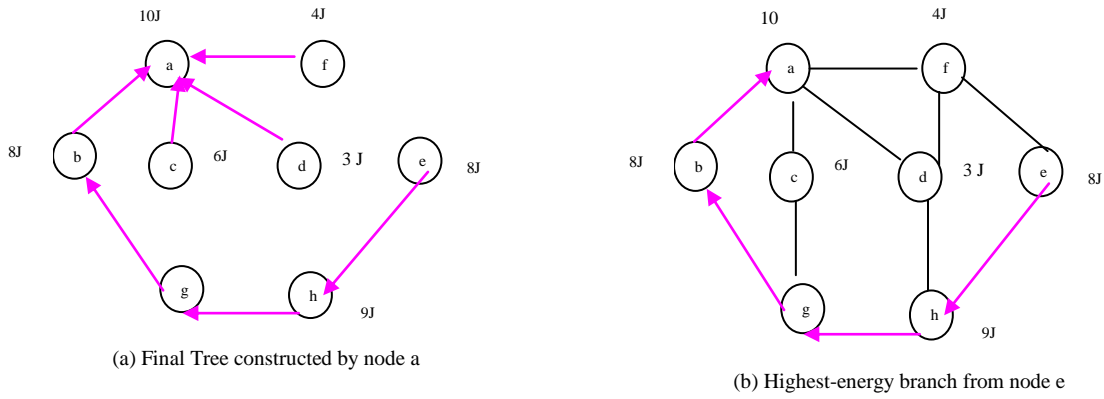

(a) Final Tree constructed by node a

(b) Highest-energy branch from node e

Fig. 1a. Final Tree construction for node a

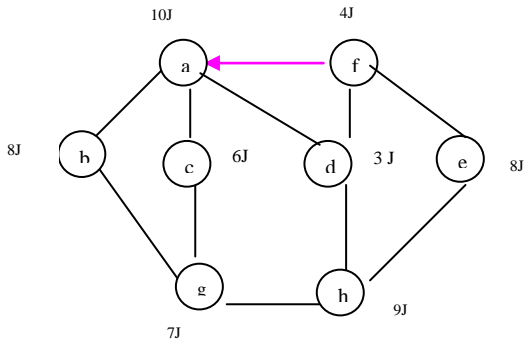

(c) Highest-energy branch from node 'f'

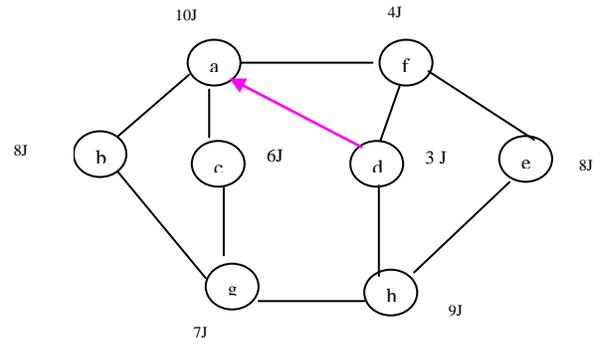

(d) Highest-energy branch from node 'd'

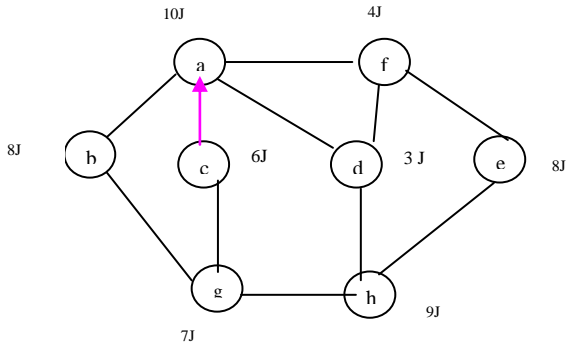

(e) Highest-energy branch from node c

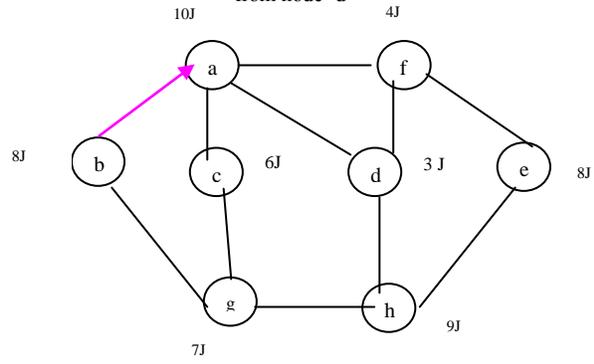

(f) Highest-energy branch from node 'b'

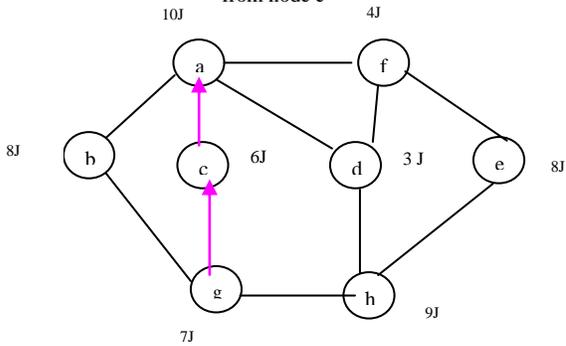

(g) Lower-energy branch from node g

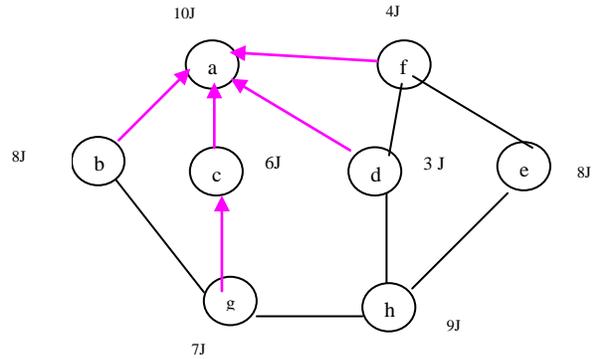

(h) Tree constructed by node a

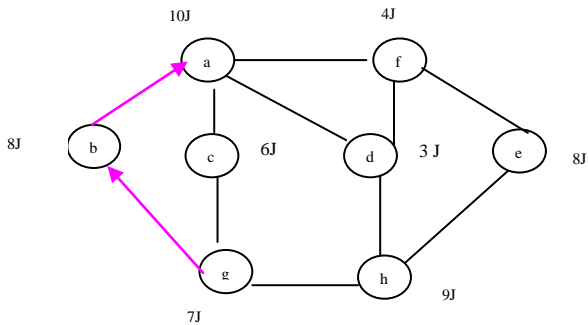

(i) Highest-energy branch from node g

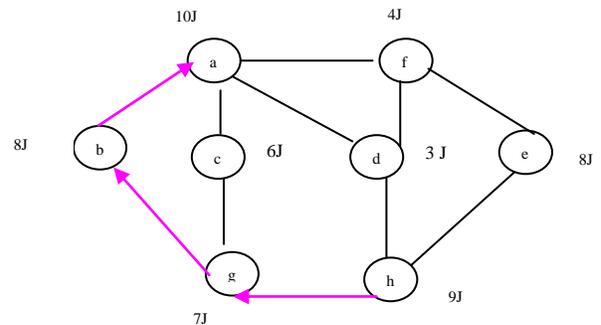

(j) Highest-energy branch from node 'h'

Fig. 1 b Exploring the highest energy branch from every source to any tree root.

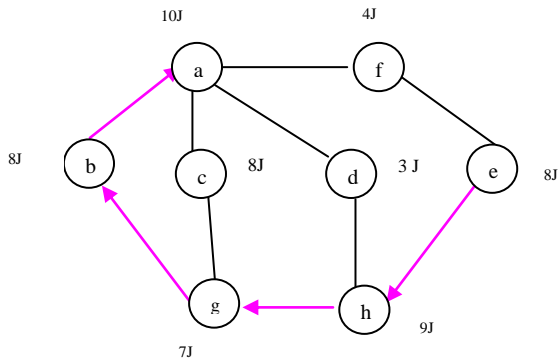

(a) Branch from node 'e'

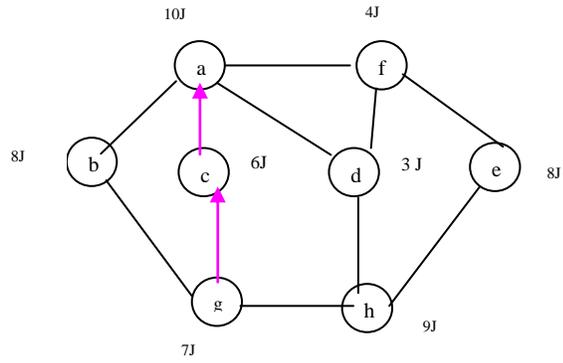

(b) Branch from node 'g'

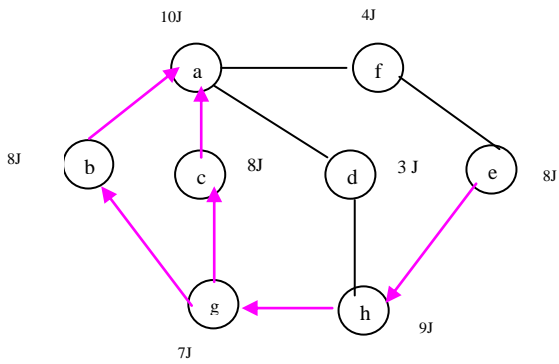

(c) Loop Detected

Fig. 2 Loop created at node g

To avoid loops we define a ***Noloop*** function .This function will take the branch as an input and will test whether loop will be created after its attachment to the tree.

---

***NoLoop*** ( node ID i , branch $brlist_{a,b,M}$ )

1. remove $eid_a$ from $brlist_{a,b,M}$ to get new $brlist_{c,b,p}$
2. where c is the node at the tail of $brlist_{c,b,p}$ and p ε $P_{c,b}$
3. if number of eids in $brlist_{c,b,p}$ < 2 or ( $brlist_{c,b,p}$ ‖ $eid_i$ ) = $brlist_{c,i,p}$(stored at $tree_i$ )
4. return true

---

Line 1and 2 removes the message initiator of node a so that the branch only contains a list of its parents. Line 3 ensures the already attached branch for node M stored in $tree_i$ matches with the route through which the received branch travels . The function always accept a branch of size less than 3 since a loop can only be created by adding a branch of size greater than 2 to its existing tree.

## C. Searching a Decentralized Lifetime Minimizing Tree For Every Source

Since each source n carries its unique tree structure stored in $tree_n$, the protocol requires every source to broadcast its tree and select the one with the highest tree energy for application independent data aggregation. Our objective is to create a tree rooted at node h shown in figure such that :

$$TE_h \geq TE_n \quad , \quad \forall n \in N , n \neq h \qquad (6)$$

$TE_h$ and $TE_n$ can be calculated using (5). There can be multiple trees that yield same tree energy . To break such ties some other properties such as tree depth , root energy and root id should be compared . To illustrate the descriptions consider Fig. 3a and Fig. 3b . By having each node to broadcast its selection there will be 8 trees under comparison . Among all theses trees , trees constructed by node b,g,e,h and a comprise the highest tree energy of 7joules. We select the tree created by node g as our lifetime minimizing tree and node g as our root for data aggregation. Since this tree has a lower tree depth than that by nodes e and a and a higher root energy yhan that by nodes b and h. We define $DLMT_a$ to be the lifetime minimizing tree that a node a currently selects and $DLMTE_a$ be the energy of $DLMT_a$ . Initially $DLMT_i$ is equal to $tree_i$ for all sources i . *Search DLMT* function described below describes the procedures to search a decentralized lifetime minimizing tree for each source .

---

***Search DLMT*** ( node $ID_i$ )

1      single-hop broadcast $DLMT_i$
2      while receiving $DLMT_a$ from node a (a $\neq$ i )
3         if *BestTree* ($DLMT_a$) ,
4            delete $DLMT_i$ and copy $DLMT_a$ to $DLMT_i$
5            calculate $DLMTE_i$
6            single-hop broadcast $DLMT_i$

---

***BestTree*** ($DLMT_i$ , $DLMT_j$)

1      if rows in ( $DLMT_j$ ) > rows in ( $DLMT_i$ )
2         return true
3      if [ rows in ($DLMT_j$) = rows in ( $DLMT_i$ ) ] and [ $DMTE_j$ > $DLMTE_i$ ]
4         return true
5      if [ rows in ( $DLMT_j$ ) = rows in ( $DLMT_i$ ) ] and [ $DMTE_j$ = $DLMTE_i$ ] and [ tree depth of $DLMT_j$ < tree depth of $DLMT_i$ ]
6         return true
7      if [ rows in ( $DLMT_j$ ) = rows in ( $DLMT_i$ ) ] and [ $DMTE_j$ = $DLMTE_i$ ] and [ tree depth of $DLMT_j$ > tree depth of $DLMT_i$ ] and [ $e_j$ > $e_i$ ]
8         return true
9      if [ rows in( $DLMT_j$ ) = rows in( $DLMT_i$ ) ] and [ $DMTE_j$ = $DLMTE_i$ ] and [tree depth of $DLMT_j$ > tree depth of $DLMT_i$ ] and [$e_j$ = $e_i$] and j < i
10     return true
11     else return false .

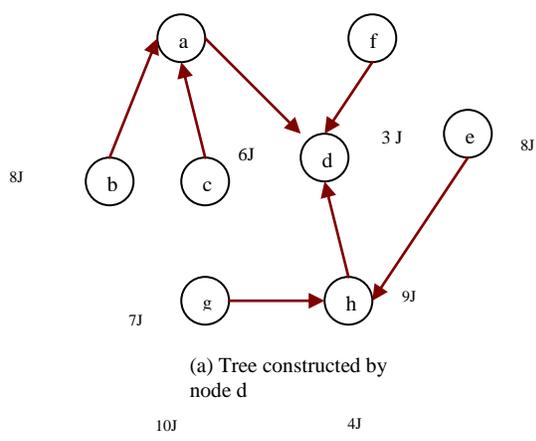

(a) Tree constructed by node d

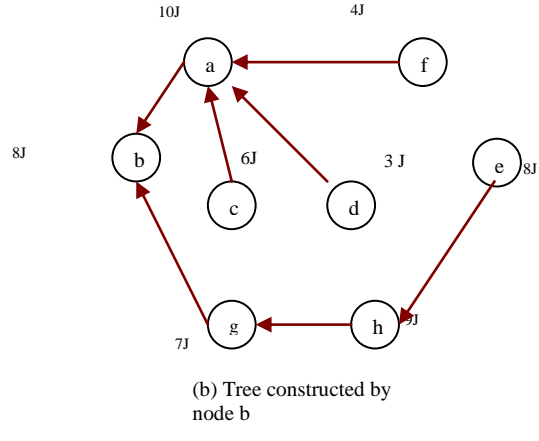

(b) Tree constructed by node b

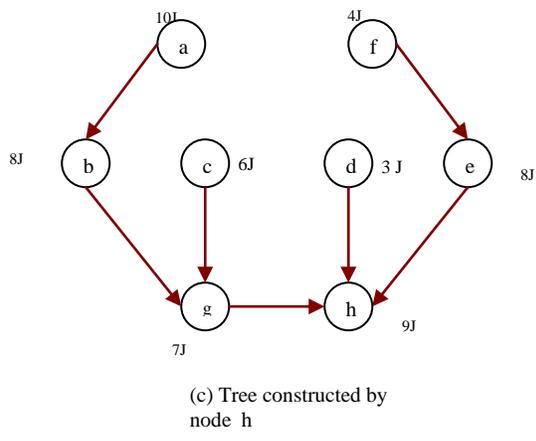

(c) Tree constructed by node h

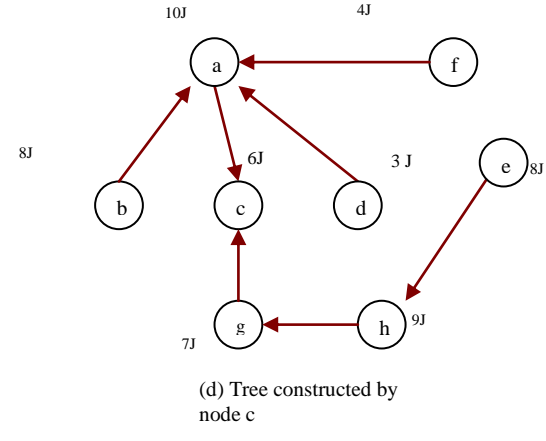

(d) Tree constructed by node c

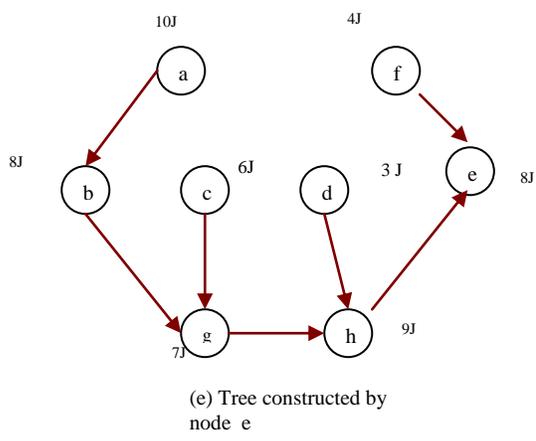

(e) Tree constructed by node e

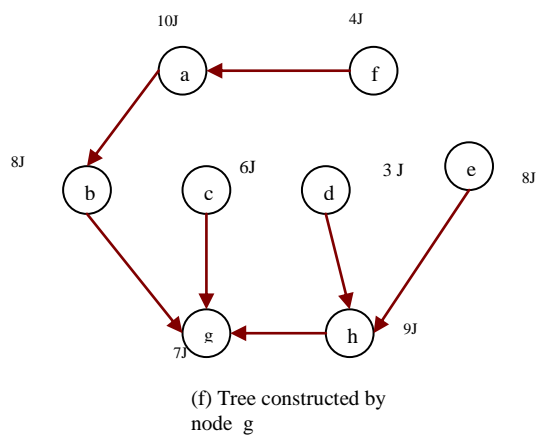

(f) Tree constructed by node g

Fig. 3 a Tree construction for various nodes to select node for data aggregation

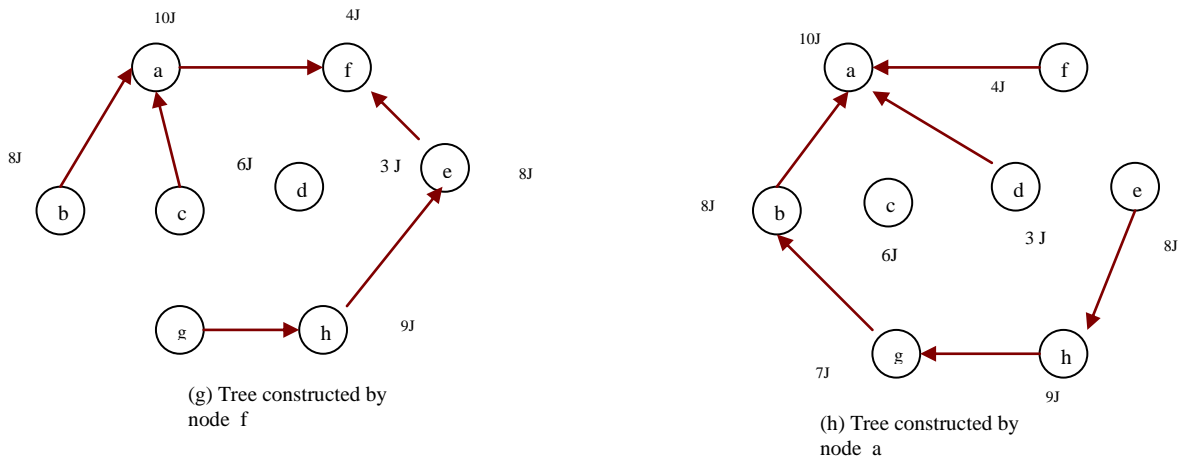

(g) Tree constructed by node f

(h) Tree constructed by node a

Fig. 3 b Tree construction for various nodes to select node for data aggregation

The *search DLMT* function searches the decentralized lifetime minimizing tree for each source .The function is applied on node i. The *BestTree* function receives DLMT tree as an input and returns true it it has more entries or if its tree energy is greater than that of what node i is currently having. Tree depth, root energy and then root ID are used to break the ties whenever necessary. $DLMTE_j$ is the energy of tree $DLMT_j$ which is calculated using (5). Tree depth of tree $DLMT_j$ is the maximum number of eids of all branches stored in $DLMT_j$.

*D. Implementation of DLMT algorithm in practical WSNs*

After explaining the algorithm of DLMT , we now proceed to define the control packet structure of DLMT for the implementation related issues. A control packet should only consists of brlist and DLMT for the ExploreBranch and Search DLMT function. Failure to receive the brlist can lead to a tree that does not have the highest tree energy . Control message $Cm_j$ should have the format [$restart_j$ , $tree_j$ , $DLMT_j$ ] . In the above format $tree_j$ and $DNLMT_j$ are two different trees , $tree_j$ is the tree created by node j whereas $DLMT_j$ is the network lifetime minimizing tree selected by node j . Forwarded Directed Diffusion (FDD) is considered as the routing platform .Hence all the control messages are wrapped with the FDD packet structure . Figure describes the packet structure of the DLMT control message . The 24-Byte header contains the information such as destination ID , source ID , packet number and packet length .The 12-byte scope and type attributes control the packets processed by the Diffusion protocol inside each node .The 12-byte control type attributes describes the type of the control message that are being exchanged in the network . Finally the 12-byte plus a space of variable size are allocated to encode all the DNLMT control information for any particular network node i.

| 24–Byte Directed Diffusion Header | |
|---|---|
| 12–Byte Directed Diffusion scope attribute | 12–Byte Directed Diffusion type attribute |
| 12-Byte control type attributes | 12-byte plus a space of variable size |

Fig. 4 Packet structure for DLMT control message

To ensure that the data from each source is always arriving at the root via its parent , a mechanism is required to construct another tree whenever parent runs out of its energy. The root should periodically broadcast a hello message to ensure this functioning. All other nodes on receiving this message from its parent should simply broadcast its message to the network. Only upstream connection towards the parent should be scanned since direction towards parent is an important issue. A timer Tm scans the connectivity with its parents periodically ,the timer Tm expires after every T seconds .A typical value of T is 25 seconds . Whenever a node losses its connection with its parent the restart flag in its control message is set to value 1 . Any other nodes that receive this flag with the value 0 will have to start the process. In this whenever a node with restart flag 1 is received the entire process is restarted and a new tree is reconstructed to build up the broken link.

*E. Algorithm for proposed Decentralized Lifetime Minimizing Tree*

Let $H_i$ be the hello message sent by node i and $t_{rec,i}$ be the time node that was the last to receive the hello message from its parent .Let $j_i$ denote the set of initiators in $tree_j$ .Algorithm described below summarizes our proposed algorithm. Single-hop broadcast refers to the operation of sending a control packet to all single-hop neighbors . Lines 1 and 2 restrict the messages to be exchanged within the event area. Line 3 broadcasts the initial control message and starts the maintenance timer. Line 4 creates an infinite loop. Lines 5 and 6, on the other hand, update the maintenance timer whenever a message is received. Since a node does not know where its parent is during the initial tree constructions, the node simply refreshes this timer upon receiving this message. Line 7 resets a node and restarts another round of tree constructions upon receiving a restart flag. Note that this line is only processed when a broken upstream link is detected by the transmitting node. Lines 8 to 18 correspond to the ***Highest Energy Branch*** function described Section III.A . Each node scans the brLists discovered by the transmitting node and updates its corresponding table entry if a scanned brList is new or carries higher branch energy. Note that the ***NoLoop*** function described above in Section III.B is used in line 9 to ensure that the attachment of any scanned brList to the existing tree structure of the receiving node does not create a loop. The tree energy is also updated in both lines 13 and 18 by using Equation 5 . Lines 19 to 27 correspond to the ***SearchDLMT*** function described in Section III.C One major difference is the additional comparison between the DLMT and the tree that the receiving node has just updated (lines 19 to 21). The reason is to ensure that a node selects is always better than the tree it has created even after an update. Lines 22 to 24 replace the DMLT with that from the transmitting node if the latter is better. Lines 25 to 27 update and broadcast the control message if a change is applied. Lines 28 to 36 correspond to the tree maintenance. Particularly, lines 30 and 31 broadcast the hello message if the node is a root when the maintenance timer expires. Lines 32 and 33 reset a node and inform all neighbors when the node loses connection with its parent. Finally, lines 34 to 36 allow a non-root node to transmit its hello message upon receiving that from its parent. Our algorithm has a complexity of **O(N2)**. Since a node needs to scan at most N brLists upon receiving any control message (line 8) and to search at most N table entries in either updating TE (tree energy ) or DMLTE (lines 13, 18, 19, 21, 22, and 24). All other lines can be executed in a constant number of iterations.

**Algorithm for proposed Decentralized Lifetime Minimizing Tree**

Initialize : reset $tree_i$ and $DLMT_i$ $\forall i \in N$

Create and append $brlist_{i,i,-}$ to tree $tree_i$ and $DLMT_i$ $\forall i \in N$

Set $restart_i$ to be true $\forall i \in N$

**Distributed LMT( node ID i , node energy $e_i$ , time t , timeframe Tf )**

```
1    if i is not an event source ,
2        return
3    else { single-hop broadcast m_i and start timer Tm that expires in T sec }
4    while true ,
5        if receiving a control message m_b from node b ,
6            restart timer Tm
7            re-initialize if restart_b is true and restart_i is false
8            for each brlist_{a,b,M} in tree_b with a ∈ B_b and M ∈ P_{a,b} ,
9                if NoLoop ( brlist_{a,b,M} ) ,
10                   if initiator a of brlist_{a,b,M} not found in tree_i ,
11                       append eid_i to brlist_{a,b,M} and update brE_{a,i,p}
```
$$\left( p \in P_{a,i} \right)$$
```
12                   add brlist_{a,i,p} and brE_{a,i,p} to tree_i
13                   calculate TE_i
14               else if min {e_i , brE_{a,b,M}} > brE_{a,i,q} (q ∈ P_{a,i}) ,
15                   remove brlist_{a,i,q} and brE_{a,i,q} from tree_i
16                   append eid_i to brlist_{a,b,M} and update brE_{a,i,p}
```
$$\left( p \in P_{a,i} \right)$$
```
17                   add brlist_{a,i,p} and brE_{a,i,p} to tree_i
18                   calculate TE_i
19           if BestTree ( tree_i ) is applied
20               delete DLMT_i and copy DLMT_x to DLMT_i
21               calculate DLMTE_i
22           if BestTree (DLMT_x)
23               delete DLMT_i and copy DLMT_x to DLMT_i
24               calculate TE_i
25           if change applied to tree_i or DLMT_i
26               set t_{rec,i} to be t
27               update and single-hop broadcast m_i
28           if timer Tm expires ,
29               set restart_i to be false
30               if i is the root of the tree in DLMT_i
31                   single-hop broadcast H_i
32               else if t > t_{rec,i} + Tf ,
33                   reinitialize and single-hop broadcast m_i
34           if receiving a hello message H_b from node b ,
35               if b is the parent of i in DLMT_i ,
36                   set t_{rec,i} to be t and single-hop broadcast H_i
```

IV. SUMMARY

To meet the demands where raw data readings are usually aggregated along their ways to be gathered at a single source prior to transmissions to any interested sink, we have proposed in this paper a Decentralized Lifetime-Minimizing Tree construction algorithm for future wireless sensor networks. The tree features in such a way that nodes with higher energy are tend to be chosen as data aggregating parents, whenever possible, so that the time to refresh this tree is extended and therefore less energy are involved in the tree maintenance. In addition, by constructing the tree in such a way, the protocol is able to lower the amount of data lost due to broken tree links before the tree reconstructions. Another attractive feature of the protocol is that the tree is most-likely to be centered in the middle of the event area, thereby reducing the delay during data collection . The conventional spanning tree fails to consider residual energy of nodes in the tree constructions. There is thus a good possibility that a high energy node can be used to forward data for some other nodes, thereby reducing its node lifetime and fastening the energy-depletion of any subsequent event source. To shorten the time and minimize the energy cost to tree reconstructions, preserves the functional lifetime of all sources. This proposed DLMT construction algorithm arranges all nodes in a way that each parent will have the maximal-available energy resources to receive data from all of its children. Such arrangement extends the time to refresh the tree and lowers the amount of data lost due to a broken tree link before the tree reconstructions.

## V. FUTURE WORK

In future we will simulate and compare our DLMT with already existing modules such as E-Spam and Forwarded Directed Diffusion Dissemination (FDDD) and validate our proposed DLMT construction algorithm.